# Interpolating moving least-squares methods for fitting potential energy surfaces: The dependence on coordinate systems for six-dimensional applications

Akio Kawano[1,a] and Gia G. Maisuradze[2,b]

[1]Holistic Simulation Research Program, the Earth Simulator Center, JAMSTEC
3173-25 Showa-machi, Kanazawa-ku, Yokohama, 236-0001, Japan

[2]Department of Chemistry and Chemical Physics Program, University of Nevada,
Reno, NV 89557, USA

Abstract

The basic formal and numerical aspects of different degree interpolated moving least-squares (IMLS) methods are studied using sixteen different combinations of coordinate system for fitting and weight functions. For the application we use six-dimensional potential energy surface (PES) of hydrogen peroxide, for which an analytic ("exact") potential is available in the literature. We systematically examine the effect of parameters in the weight function, the degree of the IMLS fit, and number of *ab initio* points. From these studies we discovered that the IMLS for almost all pairs of coordinate system show qualitatively similar behavior, however, the accuracy of the fits is noticeably different. We also found compact and accurate representations of potentials for presented degrees of IMLS.

---

a E-mail: kawanoa@jamstec.go.jp

b E-mail: gia@chem.unr.edu

1. Introduction

Potential energy surface (PES) of the molecular system plays a critical role in theoretical studies of chemical reaction dynamics. The enhanced computer facilities allow for straightforward use of *ab initio* forces in molecular dynamics simulations. It is fairly routine now to perform high-quality *ab initio* calculations for hundreds to thousands of geometries. However, the computational expenses increase rapidly with the enlargement of size of system, and also the levels of theory used in these calculations are often inadequate for the reactions. Although the results obtained by means of high-level quantum chemistry calculations are characterized with high accuracy, the calculation of high-level *ab initio* forces needed for the trajectory integrations is not always feasible. All these difficulties, which will be solved in time, prevent the global use of direct dynamics calculations, and motivate to develop the alternative approaches. Such efficient alternatives are fitting methods.

Since the 1970s a variety of PES fitting methods have been developed.[1] For example, least-square fitting methods, spline methods, reproducing Kernel Hilbert space interpolation methods, Morse-spline and rotated Morse-spline methods. These methods can be categorized as weighted or unweighted. During the past decade the local fitting method introduced by Ischtwan and Collins,[2] which is based on modified Shepard interpolation, has become widely accepted. A serious problem with unmodified Shepard method, referred to as the "flat-spot" phenomenon, is that the derivative of the interpolant is zero at every data point. This prevents the straightforward use of the Shepard method. However, this problem can be avoided by using a Taylor expansion that includes the first and second derivatives at each data points.[3] An attractive feature of the modified Shepard approach is its simplicity. It can be coupled with dynamics simulations to bias the fit, but the need for gradients and Hessians required in this method are not always available in higher-level *ab initio* calculations.

Recently, we have introduced interpolating moving least-squares methods (IMLS),[4-8] for fitting PESs. The IMLS is the least used method among the fitting methods. Three decades ago McLain[9] studied two-dimensional fits for some simple functions using zero-, first-, second-, third-, and fourth-degree IMLS methods. Recently, Ishida and Schatz[10,11] presented a scheme in which a second-degree IMLS (SD-IMLS) method is combined with modified Shepard interpolation. The IMLS methods as modified Shepard belong to the group of weighted fitting methods and involve polynomials of any desired degree. The Shepard method is in fact a zero-degree IMLS method (ZD-IMLS). The IMLS methods do not need gradients and Hessians, which makes them efficient for fitting PESs obtained by high-level *ab initio* calculations. In our studies we have done detailed analysis of IMLS methods up to nine degree for one-dimensional applications,[5] and applied IMLS to three-[4] and four-atomic systems.[6-8] The character of IMLS allows us to employ successfully several different mathematical procedures, the automatic point selection,[5,8] fixed and variable cutoff radiuses,[7,8] partial reduction of cross terms,[8] which makes IMLS more efficient and accurate. Apart from mathematical testing the classical trajectories were carried out on higher degree IMLS surfaces to study reaction rates for the O-O bond breaking in hydrogen peroxide (HOOH).[7]

The present work is one of the members of series that we started two years ago, and reports the detailed studies of IMLS dependence on coordinate system. It was shown for modified Shepard method that the choice of coordinate system is important for the accuracy of fit.[2,10-14] We also have partially touched this matter in our previous studies,[6-8] which motivated us to perform more detailed analysis in this direction. This paper illustrates the basic formal and numerical aspects of first-, second-, and third-degree IMLS methods for four different coordinate systems.

In order to make the link to the previous works we use the same sets of *ab initio* data points of HOOH calculated from the analytical potential PCPSDE developed by Kuhn *et al*.[15] We selected an analytical PES to avoid the more costly procedure of calculating *ab initio* points.

An outline of this paper is as follows: Section 2 contains a brief review of IMLS method. The sampling and coordinate systems that are used in the IMLS applications are discussed in Section 3. The results are discussed in Section 4. The summary and conclusions are given in Section 5.

2. Method

The basic aspects of the IMLS method are outlined in our previous papers[4-8] and earlier standard references.[16] It will be briefly outlined here for 1D applications. The generalization to many dimensions is straightforward.

Suppose m linearly independent functions $b_j(\mathbf{X})$ (j=1,…,m) are given and defined on the whole surface, the fitted surface is then a linear combination of these monomials $b_j$,

$$u(\mathbf{X}) = \sum_{j=1}^{m} a_j(\mathbf{X}) b_j(\mathbf{X}), \qquad (1)$$

where $a_j(\mathbf{X})$ are the coefficients. The fit of function u(**X**) to the data values $f_1,\ldots,f_N$ can be evaluated by means of error functional

$$E(u) = \sum_{i=1}^{N} w_i(\mathbf{X})[u(\mathbf{X}) - f(i)]^2 \qquad (2)$$

It is assumed that m ≤ N and we have introduced $w_i(\mathbf{X})$ a weight functions, which depend not only on data points **X**(i) as usual weight functions, they also are functions of **X**, the location on the PES where a fit is required. The weight functions have relatively large values for **X**(i) close to **X**, and relatively small for the more distant **X**(i). They can take the form of

$$w_i = \frac{1}{\|X - X(i)\|^{2n} + \varepsilon} \qquad (3)$$

where n is a small positive integer and ε is a very small positive real value that forces $w_i$ to be finite at **X** = **X**(i). We will discuss weight functions for different coordinates in Section 3B.

The coefficients $a_j(\mathbf{X})$ can be obtained from the standard formulation of the normal equations for least-squares fitting:

$$B^T W(\mathbf{X}) B \mathbf{a}(\mathbf{X}) = B^T W(\mathbf{X}) \mathbf{f}, \qquad (4)$$

where **a** and **f** are column vectors, B is a N×m matrix with elements $B_{ij} = b_i(\mathbf{X}(j))$, $B^T$ is transpose matrix, and W(**X**) is an N×N diagonal matrix whose element $W_{ii} = w_i(\mathbf{X})$. The solution **a**(**X**) to Eq. (4) provides the coefficients to the u at point **X**. Since the matrix B, known as a Vandermonde matrix, is ill-conditioned,[16] this approach can be unstable for higher-degree IMLS. In order to improve the numerical stability we use either singular value decomposition (SVD) or QR factorization approaches.

3. Characterization of sampling and coordinate systems

A. Sampling

The analytical PES allow us critically assess the accuracy of the IMLS fits. First of all, we determine two different sets of data points: set 1 is the data points used for fitting, and set 2 is a much larger set of points used for the evaluation of global rms error of the fits.

The sampling scheme plays an important role in the efficiency of fitting multidimensional PESs. In our previous study[8] we have tested four different sampling methods for fitting. It was shown that the grid method is the most efficient method to sample the data points. In present work we employ the grid sampling method. As in previous paper, we augment 89 symmetry distinct predetermined data points, selected for set 1 to cover the low-energy region of the PES, with points selected in a grid built from a geometric progression:

$$\begin{bmatrix} r_1^{(n1)} \\ r_2^{(n2)} \\ \vdots \\ r_6^{(n6)} \end{bmatrix} = \begin{bmatrix} f^{n1} & & & 0 \\ & f^{n2} & & \\ & & \ddots & \\ 0 & & & f^{n6} \end{bmatrix} \cdot \begin{bmatrix} r_1^{(0)} \\ r_2^{(0)} \\ \vdots \\ r_6^{(0)} \end{bmatrix}, \quad (5)$$

where $f > 1$, each $n_i$ is an integer, $r_1$ is the $H_1$-$H_2$ distance, $r_2$ is $H_1$-$O_1$, $r_3$ is $H_1$-$O_2$, $r_4$ is $O_1$-$H_2$, $r_5$ is $H_2$-$O_2$, $r_6$ is $O_1$-$O_2$, and

$$\begin{bmatrix} r_1^{(0)} \\ r_2^{(0)} \\ r_3^{(0)} \\ r_4^{(0)} \\ r_5^{(0)} \\ r_6^{(0)} \end{bmatrix} = \begin{bmatrix} 4.45 a_0 \\ 1.82 a_0 \\ 1.82 a_0 \\ 1.82 a_0 \\ 1.82 a_0 \\ 2.75 a_0 \end{bmatrix}. \quad (6)$$

For set 2 we use only the geometric progression to select points but with values of $f$ closer to unity which produces a finer grid. The 40 192 data points used for set 2 was generated by $f = 1.1108$, comparing to six times larger set it gives minor differences (0.6% - 1.5%) in rms errors. The energies of data points in both sets are less than 100 kcal mol$^{-1}$. The both set 1 and set 2 essentially encompass all the parts of PES that are chemically interesting for HOOH→OH+OH dissociation and association reactions. Only extraordinary high energies (>100 kcal mol$^{-1}$) or exceptionally large OH+OH separations ($>6a_0$) are excluded. Thus a satisfactory rms error implies that the spectroscopy and the dissociation/association reactivity of the HOOH system can be computed with confidence.

B. Coordinate systems

We examine four coordinate systems: 1. interatomic distance coordinates (IAD); 2. reciprocal interatomic distance coordinates (RIAD); 3. valence internal coordinates (VINT); 4. reciprocal valence internal coordinates (RVINT). In the next Sections we will use the following denotations for hybrids, e.g. VINT-RIAD, RVINT-IAD ect, the first term of pairs denotes the coordinate system used for fitting and second term denotes the coordinate system used for the weight function.

It has been shown[2,6-8,10-13] that the use of hybrids of coordinate systems is sometimes more efficient than the same coordinate systems used for both fitting and weight function coordinates. The interatomic distance and reciprocal interatomic distance coordinates were used in these studies. The denotation of interatomic distances and reciprocal interatomic distances are $\mathbf{R} = (r_1, r_2, r_3, r_4, r_5, r_6)$ and $\mathbf{X} = 1/\mathbf{R}$, respectively. The expression of weight function for interatomic and reciprocal interatomic coordinates is given in Eq. (3).

The valence internal coordinates are bond lengths, bond angles, and dihedral angles. For HOOH molecule the valence internal coordinates are $\mathbf{Z} = (r_{OH}, r_{OO}, r_{OH}, \theta_1, \theta_2, \cos\tau)$, the reciprocal valence internal coordinates are $\mathbf{T} = (1/r_{OH}, 1/r_{OO}, 1/r_{OH}, \theta_1, \theta_2, \cos\tau)$. The expressions of weight functions for valence and reciprocal valence internal coordinates are following:

$$w_i = \frac{1}{\left\{ c\left[(r_1 - r_1^{(i)})^2 + (r_2 - r_2^{(i)})^2 + (r_3 - r_3^{(i)})^2\right] + a\left[(\theta_1 - \theta_1^{(i)})^2 + (\theta_2 - \theta_2^{(i)})^2\right] + (\cos\tau - \cos\tau^{(i)})^2 \right\}^n + \varepsilon}$$

$$w_i = \frac{1}{\left\{ d\left[\frac{1}{(r_1 - r_1^{(i)})^2} + \frac{1}{(r_2 - r_2^{(i)})^2} + \frac{1}{(r_3 - r_3^{(i)})^2}\right] + a\left[(\theta_1 - \theta_1^{(i)})^2 + (\theta_2 - \theta_2^{(i)})^2\right] + (\cos\tau - \cos\tau^{(i)})^2 \right\}^n + \varepsilon}$$

(7)

where $a$, $c$, and $d$ are the coefficients, which link the distances and angles. In order to find out the optimal values for these coefficients we have calculated rms error of energies as a function of $c$ and $d$ ($a = 1$ and has a unit rad$^{-2}$). The Fig. 1 (a,b) illustrates the results. The hybrid RVINT-VINT was used for determination of optimal value for $c$ coefficient, panel a, the optimal value of coefficient $d$, was defined by RVINT-RVINT combination, panel b. The change of coordinate system for fitting coordinates does not change the region of optimal values of coefficients $c$ and $d$, so all calculations shown in next Section have been performed with coefficients $c = 1$ and $d = 10$. The other parameters of weight function used in these calculations are $\varepsilon = 1\times10^{-20}$ and $n = 5$, the number of data points in set 1 is 1689, and the degree of IMLS is second (SD-IMLS).

4. Results

As described in previous sections the IMLS fit is a function of weight function parameters ε and *n,* the degree, the number of data points N. In our previous studies[5,7,8] we have examined these dependences for some coordinate systems. Since the present work encompasses the broad spectrum of coordinate systems the results of previous studies are also included.

A. Dependence on ε

The figures 2-5 demonstrate the dependence of rms error of energy on ε for first-degree IMLS (FD-IMLS), SD-IMLS, and third-degree IMLS (TD-IMLS) fits for a fixed value of *n*. The number of data points used in these calculations is 1689. Five conclusions can be drawn from these figures. First, the most of the results given in these figures are qualitatively the same, i.e. when ε is too large the accuracy of all fits is degraded, as ε decreases the rms error reaches a minimum that essentially persists for all further decreased ε. Second, the hybrids VINT-IAD and RVINT-IAD give a qualitatively different behaviour for some degrees of IMLS (figures 4 and 5), which automatically discards these pairs from the list of useful combinations of coordinate systems. Third, with fixed value of ε, increasing the degree of IMLS decreases the rms error. Fourth, all pairs of coordinate systems show only necessity of ε to be small enough for optimally accurate fits. The ranges for the optimal values of ε are different for different pairs of coordinate systems. The figures 2-5 show that the smallest ranges have the pairs where weight function employs RIAD coordinates, however, the use of small $\varepsilon$ (~$10^{-20}$ or less) guarantees optimal performance of the weights for any coordinate system. Fifth, the range of accuracy of the fits for different coordinate systems is quite noticeable. For FD-IMLS the difference between the rms errors of best and worst fits is 2.35 (kcal mol$^{-1}$), which is 26%. For SD- and TD-IMLS these differences are larger: 2.97 (kcal mol$^{-1}$) and 2.95 (kcal mol$^{-1}$), which are in percentage 61% and 68%, respectively. The most accurate fit for FD-IMLS is obtained by hybrid RVINT-RIAD. For SD- and TD-IMLS hybrid RIAD-VINT gives the best results. However, figures 2-5 show that several other combinations of coordinate systems also should be considered as the useful hybrids since their fits are quite accurate.

B. Dependence on *n*

The figures 6-9 illustrate the behaviour of rms error of energy for FD-, SD-, TD-IMLS as a function of *n* for a fixed value of $\varepsilon$ selected at its optimal value. The number of data points is the same as in previous sub-section A studies. Almost all hybrids of coordinate systems of FD-, SD-, TD-IMLS behave in qualitatively similar ways with optimal fits occurring for large enough values of *n*. As in previous sub-section, the behaviour of some hybrids of coordinate systems, VINT-IAD and RVINT-IAD for FD- and SD-IMLS (figures 8, 9), is qualitatively different from others. These hybrids are considered as an useless combinations and eliminated from the list. The results show that the range of optimal values of *n* for all pairs starts from 4 and persists for all further larger *n*. Also, for any given value of *n*, the fitting error improves with the degree of the IMLS fit. The similar to sub-section A, the pairs RVINT-RIAD and RIAD-VINT give the smallest rms errors for FD-IMLS and SD-, TD-IMLS, respectively. The differences between the best and worst rms errors for FD-, SD-, TD-IMLS are 2.27, 2.41, 2.65 (kcal mol$^{-1}$), which are in percentages 26%, 56% and 66%, respectively.

C. Dependence on N and on degree

The results in previous sub-sections A and B indicate that the optimal fits for almost all combinations of coordinate systems can be achieved by means of large enough values of *n* and small enough values of $\varepsilon$. The tables 1-3 show the rms fitting errors for eight different numbers of data points for FD-, SD-, and TD-IMLS. The data points in both sets are sampled by the scheme described above. For almost all combinations of coordinate system the rms error always decreases as the number of data points increases, exceptions are again VINT-IAD and RVINT-IAD hybrids for FD- and SD-IMLS. Also, increasing the degree of IMLS systematically improves the fit for each coordinate system. The results for SD- and TD-IMLS (tables 2-3) indicate that the choice of coordinate system is important for accuracy of the fit, different variations of coordinate system give noticeably different rms errors. Some hybrids give better fits for lower degree IMLS than other hybrids of higher degree IMLS. There are several pairs illustrating the most accurate fits for FD-, SD- and TD-IMLS (RIAD-IAD, RIAD-RIAD, RIAD-VINT, RVINT-RIAD,

RVINT-VINT). However, it is hard to choose the best one since there is no the combination of coordinate systems giving the least values of rms error for all listed numbers of data points. The rms errors of TD-IMLS for small numbers of data points are considerably high than rms errors of FD- and SD-IMLS. It is not surprising since the number of *ab initio* points per dimension goes as $N^6$, and in the case of 189 and 289 data points we have 2.4 to 2.6 points per dimension, which are very small numbers for TD-IMLS to show the advantage over lower degrees of IMLS.[5-8] We also have investigated the power law behaviour for all combinations of coordinate system given in tables 1-3. For FD-IMLS it varies from –0.21 to –0.34, for SD- and TD-IMLS power changes from –0.52 to –0.74 and –2.77 to –2.97, respectively. As in previous studies, the inverse powers increase with the degree of IMLS. However, we should notice the main contribution to the power of TD-IMLS comes from rms errors of first two smallest data points (189 and 289). The deletion of these terms significantly drops down the power.

5. Conclusions

We have presented the basic formal and numerical aspects of IMLS methods of different degree in the context of 6D application for sixteen different pairs of coordinate system. We have performed the detailed examination of effect of weight function parameters for all 16 combinations. From this systematic behavior, we have discovered the regions of parameter space for the weight functions that allow compact and accurate representations of potentials for FD-, SD- and TD-IMLS fits. In spite of significant difference in accuracy of the fits we have found that the behaviour of IMLS methods for the majority number of pairs is qualitatively similar. Only two hybrids for FD- and SD-IMLS behaved differently. The dependence of rms error on number of data points N illustrated that rms error converges with N in inverse power law fashion with powers that increase with the degree of IMLS fit. All calculations performed in this paper are addressed to the behaviour of rms errors of energy values. In order to carry out the trajectory calculations the information of gradients is needed. We have performed the detailed investigation of behaviour of gradients for a few different pairs of coordinate systems in our previous studies,[4,5,7,8] and found the qualitative similarities with rms error of energy. The scale of rms error of gradients is much larger, because the range of gradients is much larger than the range of energy values. Naturally, the following question can be raised: how accurate should be a fit of energy values which allows the accurate chemical dynamics. We partially studied this question in our previous work[7] for HOOH using valence internal and interatomic coordinates, and found inconsistency between rms errors of different coordinate systems and accurate rate constant for O-O bond fission. In other words, the different coordinate systems require different accuracy of the rms fits to achieve accurate dynamics properties. Thus, the complete answer on this question can be given after systematic dynamics simulations for different systems using different pairs of coordinate systems. One of these studies is in progress for five-atomic system.

Acknowledgements

The authors thank Professors Donald Thompson and David Leitner for computational resources.

Table 1. The rms errors of energy in kcal/mol of all hybrids for FD-IMLS, $\varepsilon = 1\times10^{-20}$, $n = 5$.

| Pairs | 189 | 289 | 489 | 889 | 1689 | 3289 | 6489 | 12889 |
|---|---|---|---|---|---|---|---|---|
| IAD-IAD | 18.64 | 16.13 | 13.11 | 10.06 | 8.83 | 7.43 | 6.80 | 5.72 |
| IAD-RIAD | 13.75 | 10.41 | 9.53 | 8.18 | 6.71 | 6.30 | 5.32 | 3.95 |
| IAD-VINT | 17.47 | 14.41 | 12.42 | 9.97 | 8.59 | 7.56 | 6.92 | 5.68 |
| IAD-RVINT | 12.78 | 10.74 | 10.02 | 8.81 | 7.75 | 7.14 | 6.20 | 4.72 |
| RIAD-IAD | 19.90 | 14.92 | 12.67 | 9.82 | 8.43 | 7.11 | 6.45 | 5.49 |
| RIAD-RIAD | 13.27 | 10.76 | 9.69 | 8.18 | 6.69 | 6.20 | 5.17 | 3.95 |
| RIAD-VINT | 19.28 | 14.12 | 11.90 | 9.27 | 7.88 | 6.91 | 6.28 | 5.18 |
| RIAD-RVINT | 12.63 | 10.97 | 9.76 | 8.27 | 7.09 | 6.59 | 5.57 | 4.31 |
| VINT-IAD | 18.40 | 54.39 | 16.11 | 12.09 | 23.61 | 9.53 | 9.68 | 6.09 |
| VINT-RIAD | 11.96 | 10.50 | 9.39 | 8.12 | 6.64 | 6.13 | 5.17 | 3.77 |
| VINT-VINT | 17.32 | 14.75 | 13.02 | 10.37 | 8.91 | 7.63 | 7.07 | 5.78 |
| VINT-RVINT | 12.46 | 10.99 | 10.23 | 9.10 | 7.88 | 7.17 | 6.26 | 4.71 |
| RVINT-IAD | 19.39 | 46.85 | 13.92 | 13.95 | 17.21 | 12.38 | 11.54 | 5.56 |
| RVINT-RIAD | 12.78 | 11.00 | 9.59 | 8.13 | 6.56 | 5.98 | 4.96 | 3.72 |
| RVINT-VINT | 19.16 | 14.86 | 12.79 | 10.01 | 8.40 | 7.32 | 6.68 | 5.52 |
| RVINT-RVINT | 13.28 | 11.69 | 10.48 | 9.10 | 7.78 | 7.08 | 6.11 | 4.71 |

Table 2. The rms errors of energy in kcal/mol of all hybrids for SD-IMLS, $\varepsilon = 1 \times 10^{-20}$, $n = 5$.

| Pairs | 189 | 289 | 489 | 889 | 1689 | 3289 | 6489 | 12889 |
|---|---|---|---|---|---|---|---|---|
| IAD-IAD | 17.20 | 15.02 | 7.34 | 4.96 | 4.85 | 3.71 | 2.85 | 2.00 |
| IAD-RIAD | 15.71 | 10.32 | 5.95 | 4.52 | 3.64 | 3.12 | 2.44 | 1.57 |
| IAD-VINT | 16.88 | 9.33 | 6.45 | 4.63 | 3.56 | 3.16 | 2.57 | 1.84 |
| IAD-RVINT | 13.97 | 8.17 | 6.15 | 5.10 | 4.04 | 3.46 | 2.75 | 1.89 |
| RIAD-IAD | 7.70 | 4.75 | 3.72 | 2.53 | 2.00 | 1.60 | 1.25 | 1.02 |
| RIAD-RIAD | 7.77 | 4.79 | 3.79 | 2.93 | 2.44 | 1.88 | 1.66 | 1.31 |
| RIAD-VINT | 6.45 | 4.79 | 3.63 | 2.46 | 1.88 | 1.55 | 1.22 | 1.04 |
| RIAD-RVINT | 7.19 | 4.73 | 3.97 | 3.00 | 2.52 | 2.05 | 1.77 | 1.50 |
| VINT-IAD | 18.80 | 19.98 | 9.32 | 8.57 | 14.34 | 6.41 | 14.46 | 2.51 |
| VINT-RIAD | 13.10 | 9.85 | 6.19 | 4.57 | 3.72 | 3.19 | 2.50 | 1.50 |
| VINT-VINT | 17.69 | 11.47 | 7.06 | 4.95 | 3.94 | 3.47 | 2.67 | 1.90 |
| VINT-RVINT | 13.60 | 9.63 | 6.16 | 4.96 | 4.06 | 3.57 | 2.69 | 1.76 |
| RVINT-IAD | 9.11 | 6.32 | 3.88 | 2.95 | 2.21 | 2.25 | 1.44 | 0.86 |
| RVINT-RIAD | 8.24 | 8.19 | 3.89 | 2.94 | 2.44 | 1.95 | 1.57 | 1.01 |
| RVINT-VINT | 9.45 | 6.69 | 3.91 | 2.98 | 2.18 | 1.74 | 1.32 | 0.94 |
| RVINT-RVINT | 9.00 | 7.52 | 4.36 | 3.40 | 2.83 | 2.25 | 1.83 | 1.40 |

Table 3. The rms errors of energy in kcal/mol of all hybrids for TD-IMLS, $\varepsilon = 1\times 10^{-20}$, $n = 5$.

| Pairs | 189 | 289 | 489 | 889 | 1689 | 3289 | 6489 | 12889 |
|---|---|---|---|---|---|---|---|---|
| IAD-IAD | 86.85 | 25.26 | 6.07 | 5.21 | 3.43 | 3.33 | 1.42 | 0.96 |
| IAD-RIAD | 115.02 | 31.84 | 5.36 | 3.23 | 2.21 | 1.86 | 1.28 | 0.81 |
| IAD-VINT | 84.30 | 14.17 | 5.29 | 3.65 | 2.26 | 1.70 | 1.17 | 0.86 |
| IAD-RVINT | 118.66 | 24.16 | 7.03 | 3.90 | 3.29 | 2.23 | 1.68 | 1.03 |
| RIAD-IAD | 41.02 | 6.32 | 3.33 | 1.81 | 1.39 | 1.11 | 0.76 | 0.60 |
| RIAD-RIAD | 35.41 | 6.59 | 3.29 | 2.17 | 1.71 | 1.23 | 1.03 | 0.81 |
| RIAD-VINT | 95.19 | 3.78 | 2.92 | 1.75 | 1.37 | 1.04 | 0.79 | 0.64 |
| RIAD-RVINT | 67.14 | 4.14 | 3.24 | 2.29 | 1.81 | 1.34 | 1.14 | 0.93 |
| VINT-IAD | 112.02 | 26.46 | 8.87 | 7.59 | 4.32 | 4.72 | 3.27 | 1.31 |
| VINT-RIAD | 104.46 | 11.96 | 5.50 | 3.99 | 2.39 | 2.03 | 1.43 | 0.79 |
| VINT-VINT | 70.85 | 16.39 | 6.65 | 4.19 | 2.66 | 1.89 | 1.27 | 0.99 |
| VINT-RVINT | 102.80 | 22.63 | 6.82 | 4.57 | 4.02 | 2.61 | 2.21 | 1.26 |
| RVINT-IAD | 53.17 | 9.19 | 3.29 | 2.14 | 1.51 | 1.10 | 0.83 | 0.59 |
| RVINT-RIAD | 27.14 | 13.70 | 3.62 | 2.44 | 1.79 | 1.29 | 1.06 | 0.65 |
| RVINT-VINT | 50.06 | 8.48 | 3.33 | 2.53 | 1.54 | 1.19 | 0.79 | 0.68 |
| RVINT-RVINT | 56.12 | 9.72 | 4.14 | 2.81 | 1.99 | 1.55 | 1.22 | 0.99 |

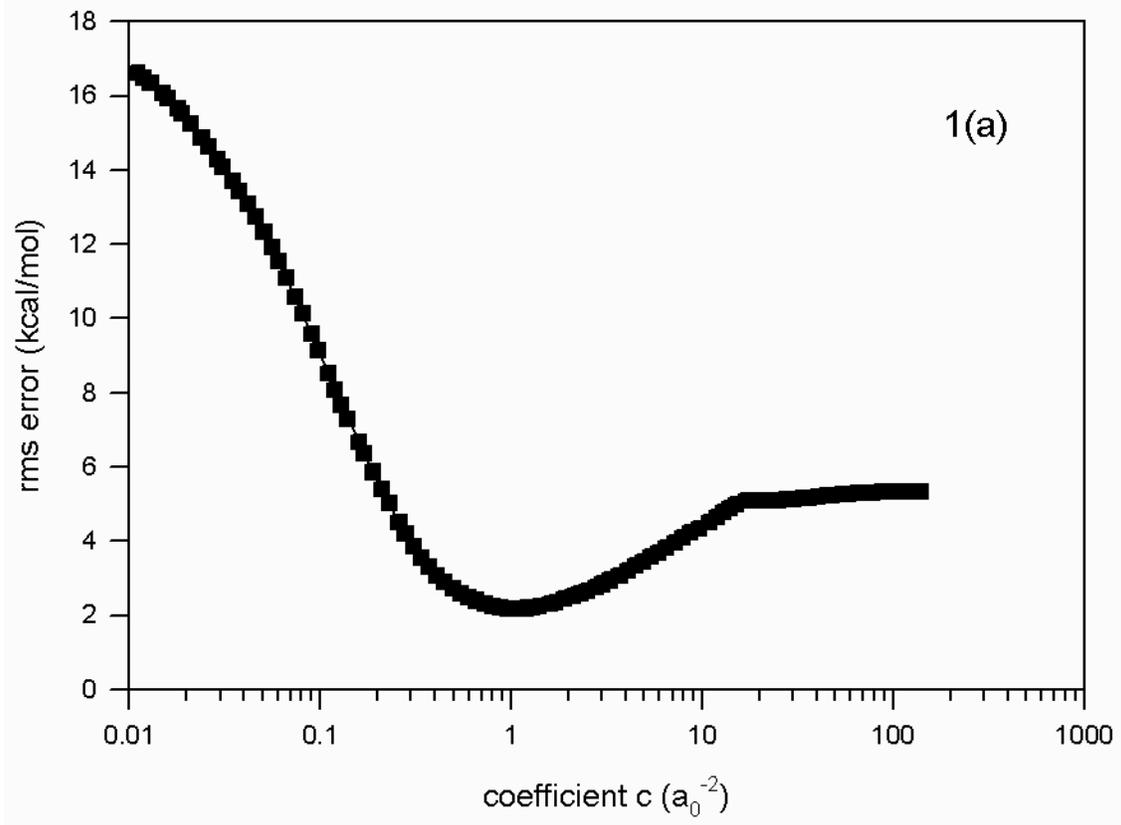

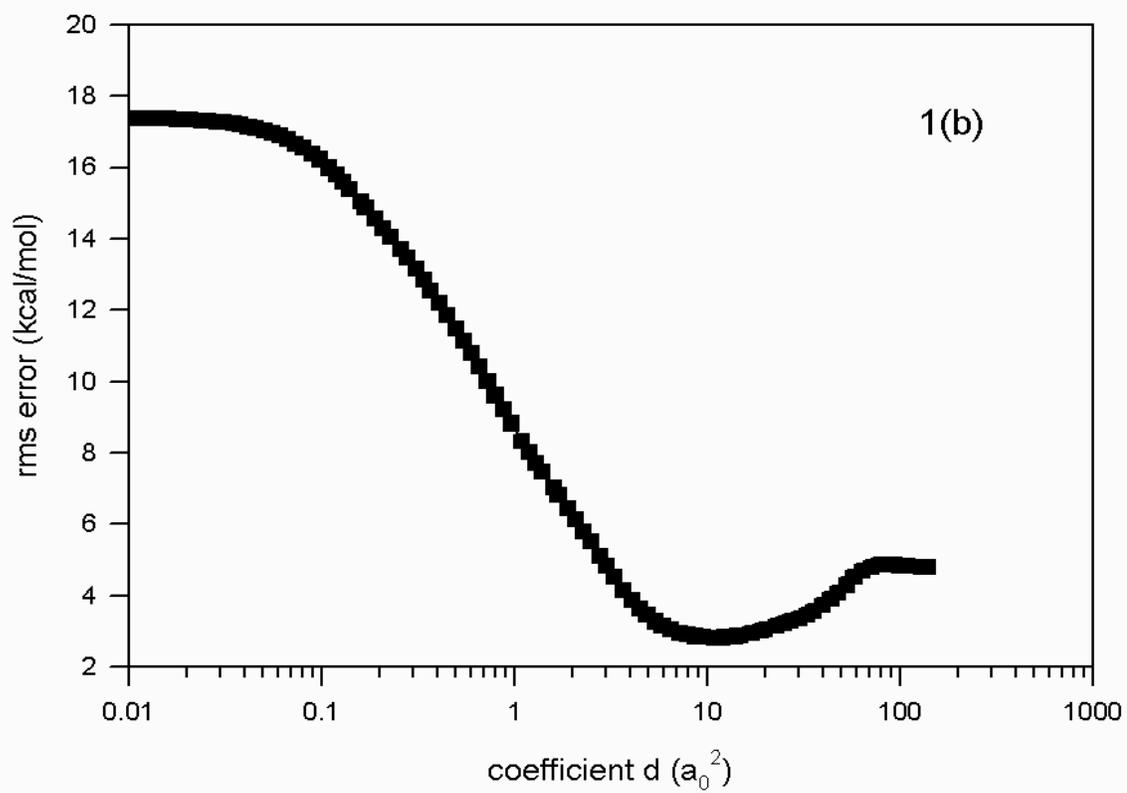

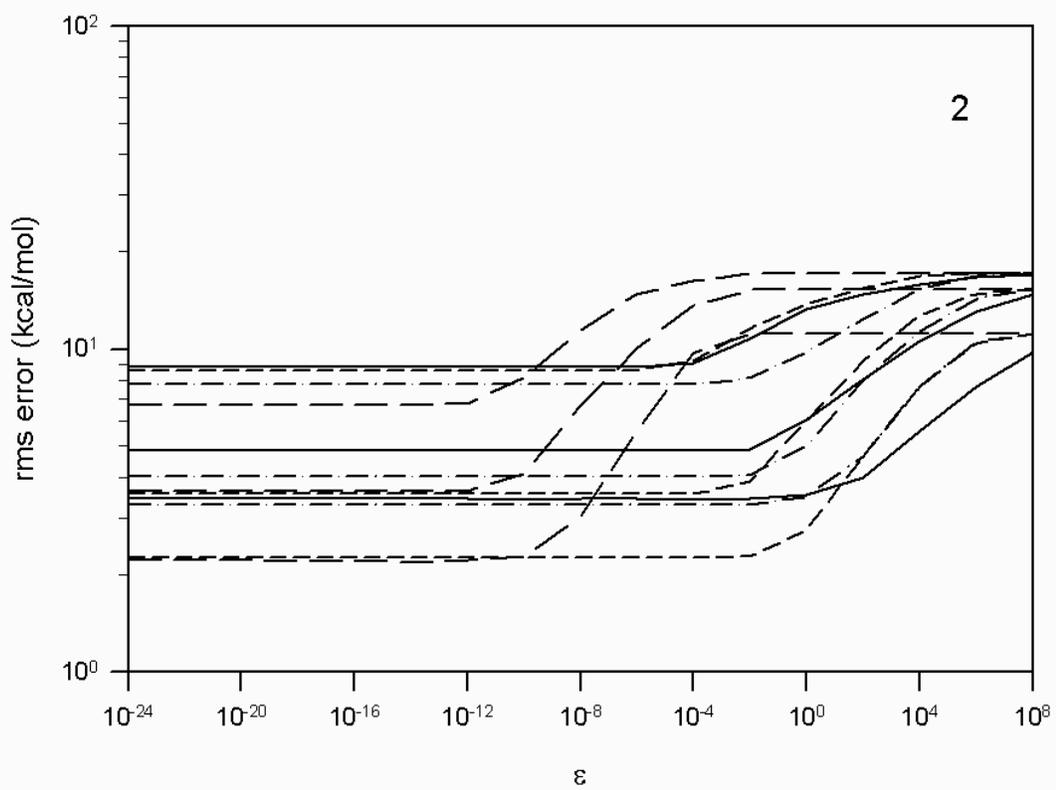

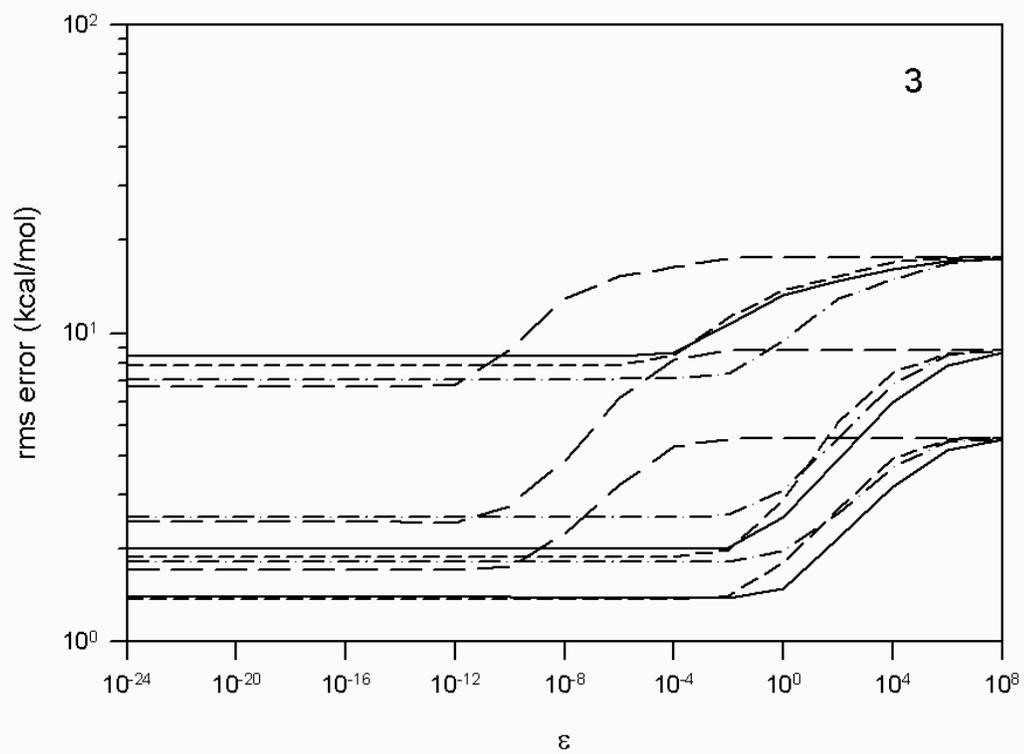

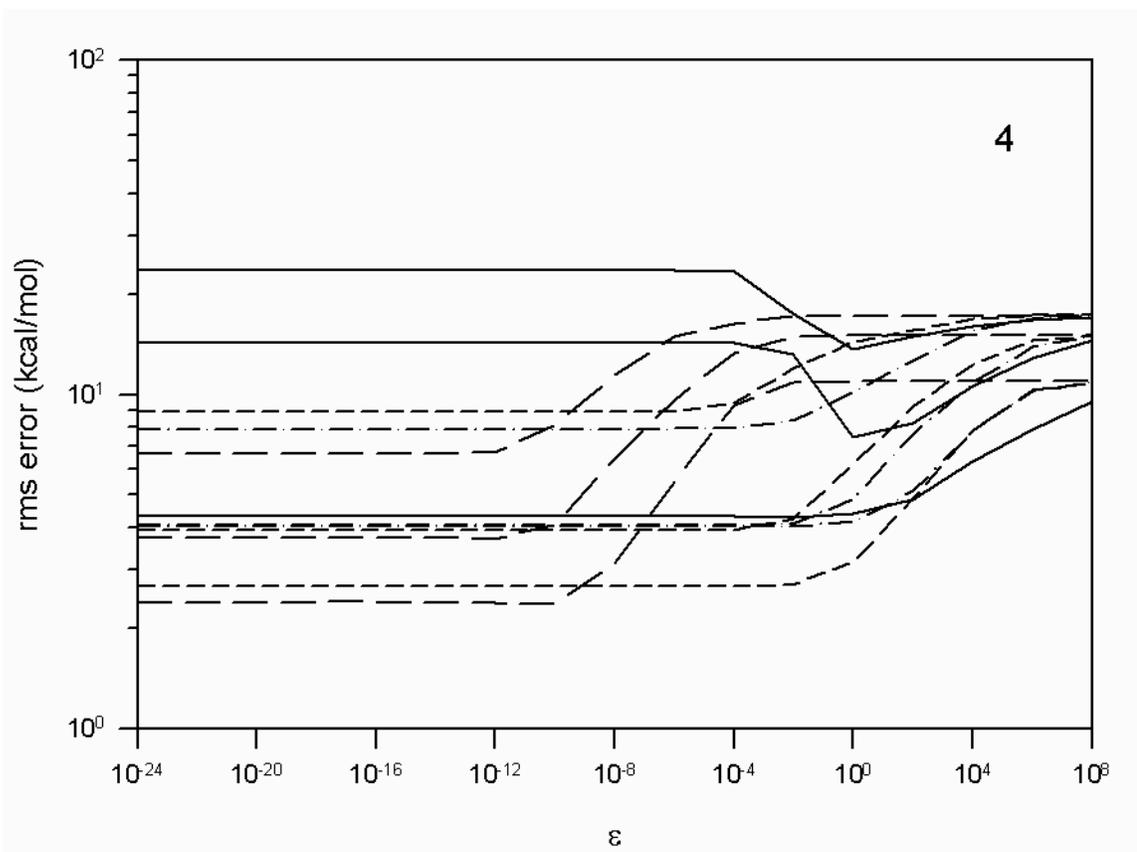

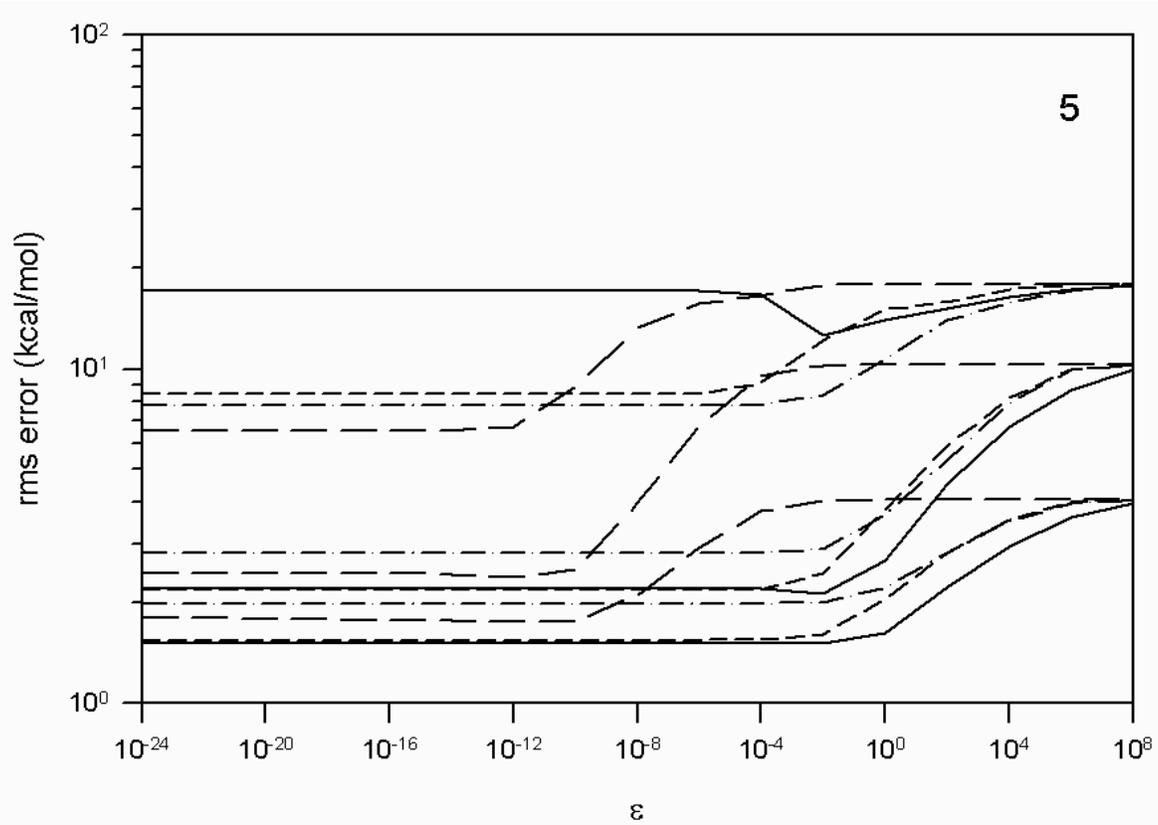

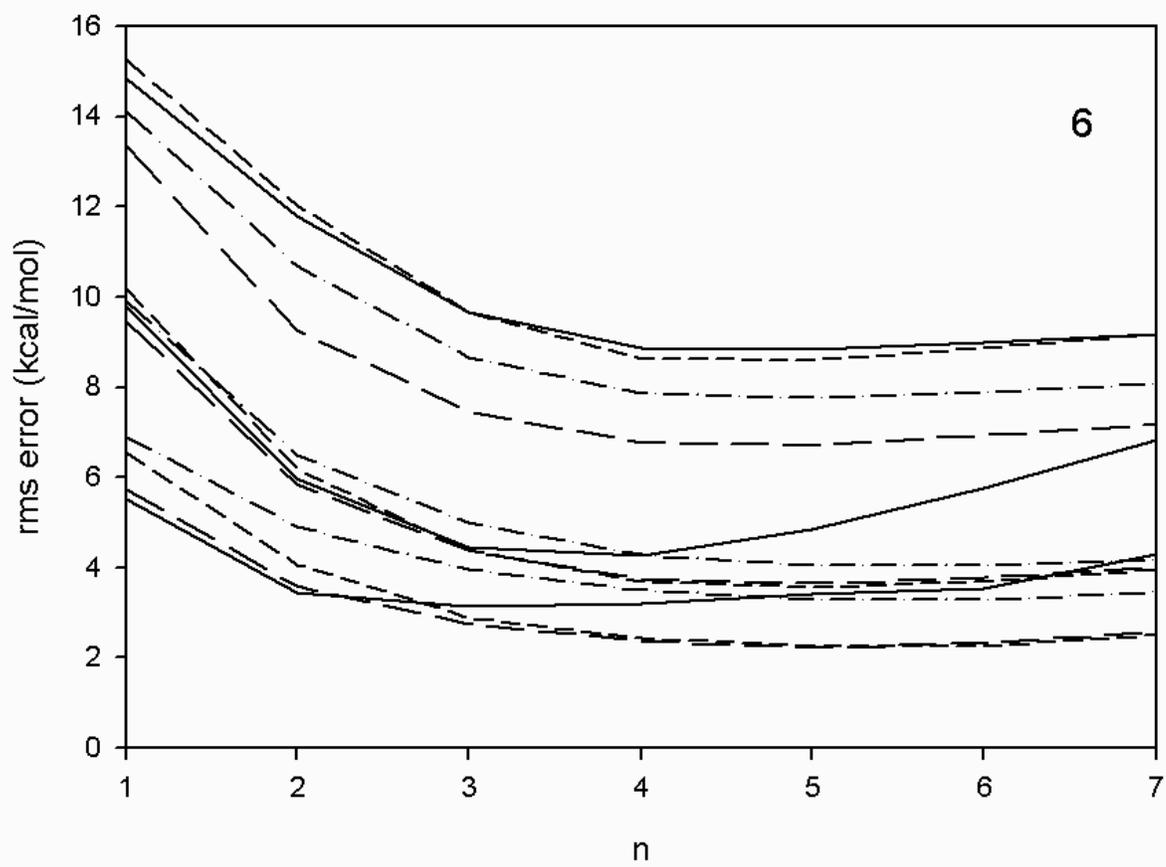

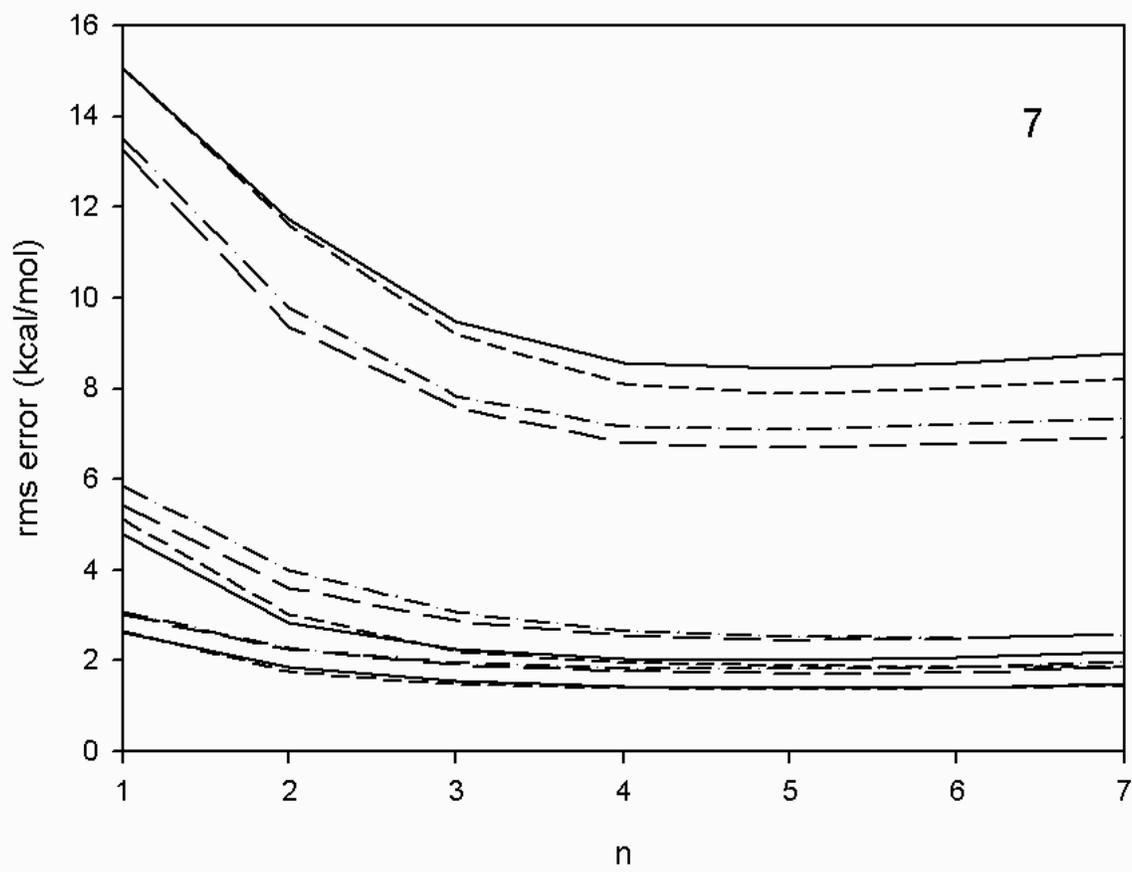

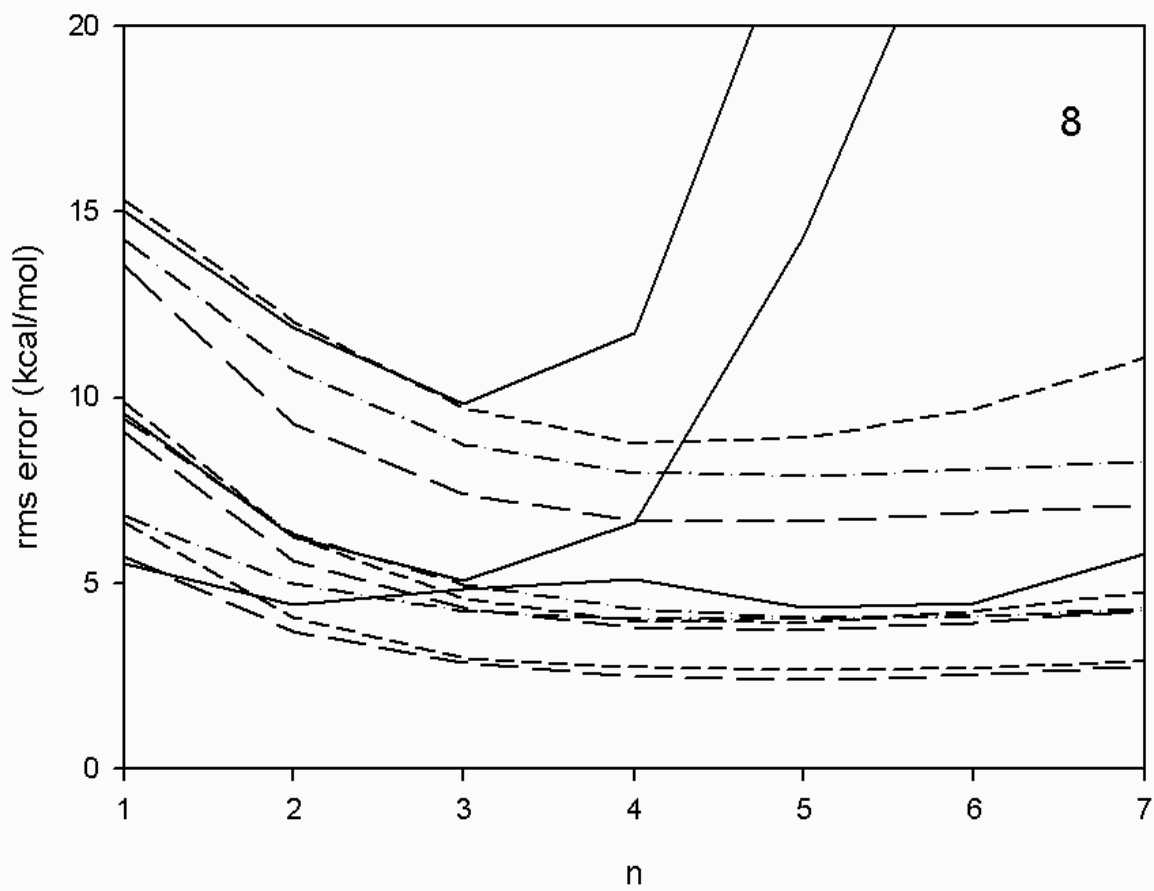

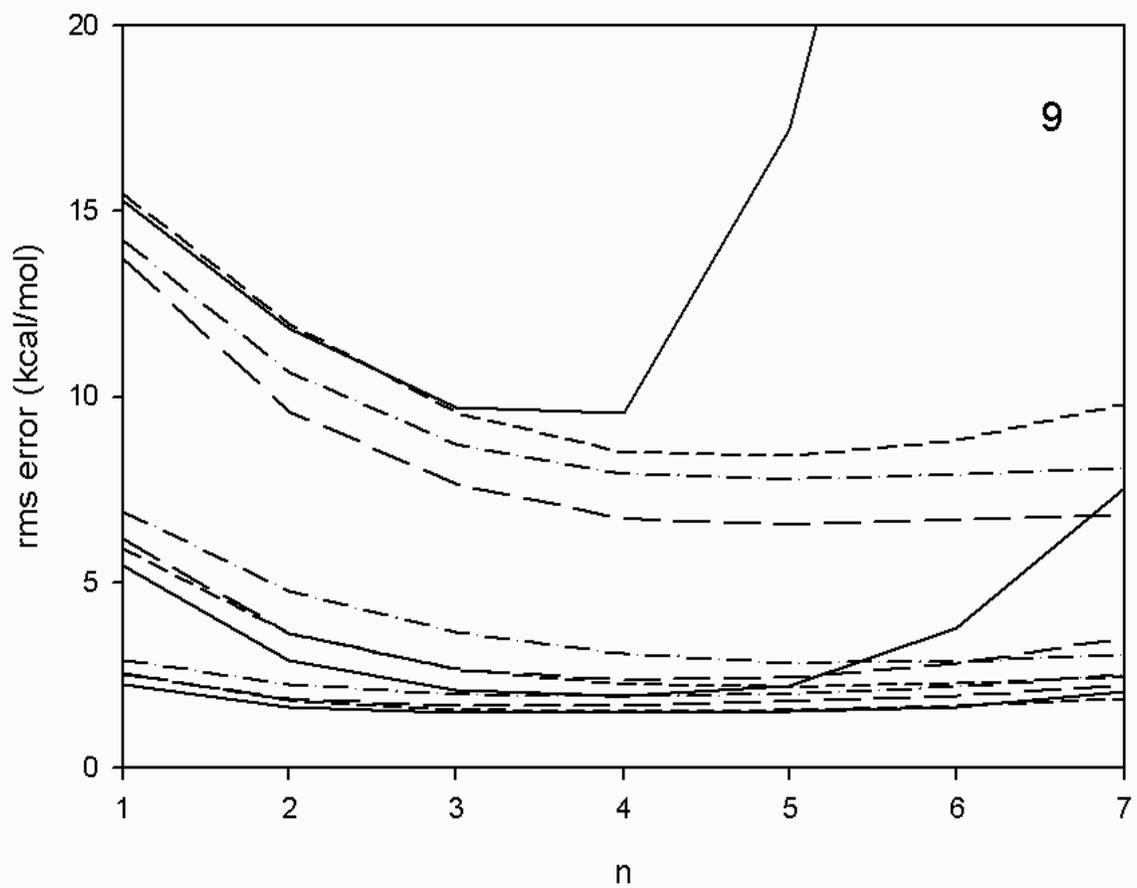

Figure Captions

Fig.1. The rms error for the potential energy in kcal/mol vs c coefficient for RVINT-VINT hybrid (a), and vs d coefficient for RVINT-RVINT combination (b). In both calculations was used SD-IMLS and $\varepsilon = 1 \times 10^{-20}$, $n = 5$, $N = 1689$.

Fig. 2. The rms error for the potential energy in kcal/mol vs $\varepsilon$ for the pairs IAD-IAD (solid), IAD-RIAD (long dash), IAD-VINT (short dash), IAD-RVINT (dash-dot). Illustrated results are for FD-, SD- and TD-IMLS. In all cases $n = 5$, N=1689.

Fig. 3. As in Fig. 2 only for pairs RIAD-IAD, RIAD-RIAD, RIAD-VINT, RIAD-RVINT.

Fig. 4. As in Fig. 2 only for pairs VINT-IAD, VINT-RIAD, VINT-VINT, VINT-RVINT.

Fig.5. As in Fig. 2 only for pairs RVINT-IAD, RVINT-RIAD, RVINT-VINT, RVINT-RVINT.

Fig. 6. The rms error for the potential energy in kcal/mol vs $n$ for the pairs IAD-IAD (solid), IAD-RIAD (long dash), IAD-VINT (short dash), IAD-RVINT (dash-dot). Illustrated results are for FD-, SD- and TD-IMLS. In all cases $\varepsilon = 1 \times 10^{-20}$, N=1689.

Fig. 7. As in Fig. 6 only for pairs RIAD-IAD, RIAD-RIAD, RIAD-VINT, RIAD-RVINT.

Fig. 8. As in Fig. 6 only for pairs VINT-IAD, VINT-RIAD, VINT-VINT, VINT-RVINT.

Fig. 9. As in Fig. 6 only for pairs RVINT-IAD, RVINT-RIAD, RVINT-VINT, RVINT-RVINT.